\documentclass[12pt, a4paper]{article}  
\usepackage{typearea}
\typearea{12}
\input{epsf.tex}
\begin{document}
\setlength{\parskip}{2ex}
\newcommand{\bi}[1]{\bibitem{#1}}
\def\be{\begin{equation}}
\def\ee{\end{equation}}
\def\bea{\begin{eqnarray}}
\def\eea{\end{eqnarray}}
\def\6{\partial} \def\a{\alpha} \def\b{\beta}
\def\g{\gamma} \def\d{\delta} \def\ve{\varepsilon} \def\e{\epsilon}
\def\z{\zeta} \def\h{\eta} \def\th{\theta}
\def\vt{\vartheta} \def\k{\kappa} \def\l{\lambda}
\def\m{\mu} \def\n{\nu} \def\x{\xi} \def\p{\pi}
\def\r{\rho} \def\s{\sigma} \def\t{\tau}
\def\Ph{\phi} \def\ph{\varphi} \def\ps{\psi}
\def\o{\omega} \def\G{\Gamma} \def\D{\Delta}
\def\Th{\Theta} 
\def\Lam{\Lambda} 
\def\S{\Sigma}
\def\PH{\Phi} \def\Ps{\Psi} \def\O{\Omega}
\def\sm{\small} \def\la{\large} \def\La{\Large}
\def\LA{\LARGE} \def\hu{\huge} \def\Hu{\Huge}
\def\ti{\tilde} \def\wti{\widetilde}
\def\non{\nonumber\\}
\def\ll{\Longleftarrow}
\def\lr{\Longrightarrow}
\def\semidirect{\;{\rlap{$\subset$}\times}\;}
\def\DA{{\buildrel {A}\over{D}}}
\def\DG{{\buildrel {\G}\over{D}}}
\def\DGl{{\buildrel {\G^{(L)}}\over{D}}}
\def\stareq{\ {\buildrel{*}\over =}\ }
\def\rG{{\buildrel {\{\}} \over \G}}
\def\rR{{\buildrel {\{\}} \over R}}
\def\rE{{\buildrel {\{\}} \over G}}
\def\xt{{\tilde x}}
\def\FF{{\cal F}}
\def\GG{{\cal G}}


\title{BRST--Antifield--Treatment of Metric--Affine Gravity}
\author{Frank Gronwald\footnote{e-mail:fg@thp.uni-koeln.de} 
\\ Institute for Theoretical Physics \\
 University of Cologne, 50923 K\"oln\\ GERMANY}
\date{}
\maketitle

{\footnotesize
\abstract{The metric--affine gauge theory of gravity provides a broad 
framework in which gauge theories of gravity can be formulated. In 
this article we fit metric--affine gravity into the covariant 
BRST--antifield formalism in order to obtain gauge fixed quantum actions. 
As an example the gauge fixing of a general two--dimensional model of 
metric--affine gravity is worked out explicitly. The result is shown to 
contain the gauge fixed action of the bosonic string in conformal gauge
as a special case.}}

\vspace*{1cm}

\vfill
{}
\pagebreak

\normalsize

\section{Introduction}

In search of a satisfactory theory of quantum gravity there still exists, 
in addition to the new methods of canonical quantization of general 
relativity, supergravity, and superstring theories, the more traditional 
approach which is based on a gauging of non--supersymmetric extensions of the 
Poincar\'e group. It developed from the early papers of Utiyama \cite{utiy56},
Sciama \cite{scia62, scia64}, and Kibble \cite{kibb61}.
The main motivation to follow this approach is provided
by the fact that the spacetime symmetry which is nowadays 
really observed is the Poincar\'e symmetry. Within the concept of a 
relativistic field theory, it is more natural to consider {\it local} 
spacetime symmetries rather than keeping spacetime symmetry {\it rigid}. 
Therefore it is natural to gauge
the Poincar\'e group. In this scheme, general relativity can be 
straightforwardly derived as a gauge theory of translations. 
However, in a perturbative expansion
the quantization of gravity models which are based on a gauging of the 
Poincar\'e group leads to either non--unitarity or 
non--renormalizability\footnote{It is 
nevertheless interesting to note that, as already observed in \cite{sezg80}, 
the presence of extra symmetries in certain models of Poincar\'e gauge 
theory \cite{bars83, blag86} might yield surprising features. The role of
extra symmetries in the context of metric--affine gravity was recently 
discussed in \cite{kalm97}.} \cite{sezg80, sezg81, kuhf86}. In order 
to `repair' these defects, it has been suggested that the 
Poincar\'e group originates from a symmetry reduction of one of its extensions
\cite{neem88}. A possible 
and fairly general framework of such a mechanism is provided by the 
metric--affine gauge theory of
gravity (MAG) \cite{hehl95}. MAG is based on a gauging of the 
$n$--dimensional affine group $A(n,R)=T^n\semidirect GL(n,R)$, i.e. the 
semidirect product of the translation group $T^n$ and the group of general 
linear transformations $GL(n,R)$. The affine group enlarges the transformations
of the Poincar\'e group by dilation and shear transformations. For a 
particular model of MAG a spontaneous symmetry breaking mechanism was 
constructed \cite{lee92}, indicating renormalizability, but with the proof of 
unitarity left as an open problem.

In this article we reconsider the approach to the quantization of MAG
and fit MAG into the covariant Becchi-Rouet-Stora-Tyutin-(BRST)-antifield 
formalism in order 
to demonstrate how to obtain gauge fixed quantum actions. 
The BRST--antifield formalism was developed by Batalin and Vilkovisky
\cite{bata81, bata83}, using earlier ideas of Zinn--Justin \cite{zinn75} 
and others \cite{dewi79}. It relies heavily on the concept 
of BRST--symmetry \cite{henn92} and was mainly developed in view of the 
quantization of gauge theories which are characterized by open or reducible 
gauge algebras. However, the antifield formalism seems 
to become a standard tool of 
quantum field theory \cite{gomi95, wein96} which is also of considerable 
use in the context of the closed and irreducible Yang--Mills theory. The more 
standard but less general Feynman--DeWitt--Faddeev--Popov method is suitable 
to covariantly gauge fix Yang--Mills theory \cite{fadd67} and general 
relativity \cite{feyn63, dewi64}. But it cannot 
straightforwardly be applied to MAG since the gauge algebra of MAG contains 
field dependent structure functions and thus constitutes no Lie--algebra.

In the literature we have found several BRST--formulations 
of specific gauge models of gravity which are included in MAG, see for example 
\cite{baul84, lee89, mori94, kalm97}. In these gauge models the generator
of translations on the spacetime manifold is taken to be a 
{\it non--gauge--covariant}
Lie--derivative. This yields a corresponding gauge algebra which is
field independent. However, a gauge--covariant notion of translational
invariance is required if matter is included which transforms 
non--trivially under the linear part of the external gauge group\footnote{The 
linear part of the external gauge group is, in 
general, the group $GL(n,R)$ or, after the symmetry reduction, the Lorentz 
group $SO(1, n-1)$.}, e.g.\ fermionic matter. Such matter can be covariantly 
translated on the spacetime manifold by the use of
a {\it gauge--covariant} Lie--derivative. The corresponding gauge algebra of 
MAG, as  already mentioned above, is field dependent, and it is this general
case we will deal with in this paper.

In order to put MAG into the BRST--antifield formalism we will proceed as 
follows: In Sec.\ 2 we will shortly review
MAG and later derive its gauge algebra in Sec.\ 3. The gauge algebra is the 
main ingredient of MAG to be inserted into the BRST--antifield formalism.
This will be done in Sec.\ 4 in order to display the BRST--symmetry 
of MAG. In Sec.\ 5 we will outline the general process of gauge fixing and 
explicitly apply it to a general two--dimensional model of MAG. This yields a
corresponding gauge fixed quantum action. The quantum action of the bosonic 
string in conformal gauge is derived as a special case from this. 

\section{MAG as a classical gauge field theory}
A physical theory constitutes a gauge theory if some of its dynamical fields 
are to be expressed with respect to a certain reference frame, the specific
choice of which is pointwise determined only modulo symmetry transformations.
These are the gauge transformations. In this case of a local symmetry 
we need to describe the equivalence of reference frames  
at different spacetime points in order to define the differential of a field.
This requires the introduction of a gauge potential, i.e. a gauge connection,
and  establishes the gauging of the symmetry group. In a physical gauge 
theory the 
gauge potential is usually made a dynamical variable, for example by adding a 
corresponding kinetic term on the Lagrangian level. 

This pattern can be followed to build up MAG: One starts from an 
$n$--dimensional, differential base manifold which represents spacetime. 
At each point $x$ on $M$  it is possible to define a 
tangent space $T_x M$, an affine tangent space $A_x M$, and an affine frame 
$(e_a, p)(x)$ \cite{koba63}.
If physical fields are to be described in affine frames, the postulate 
of local affine invariance requires the introduction of an affine gauge 
connection $(\G^{(T)\a}, \G_\a{}^\b)$. Here, $\G^{(T)\a}$ denotes the 
translational part of the affine gauge connection which accounts for local 
translation invariance while $\G_\a{}^\b$ denotes the linear part of the 
affine gauge connection which accounts for local $GL(n,R)$--invariance.
This completes the gauging of the affine group.

To arrive at a gravity theory we next have to turn the affine gauge 
connection $(\G^{(T)\a}, \G_\a{}^\b)$ into an external spacetime structure.
This step has no analogue in the case of an internal gauge theory, such as 
ordinary Yang--Mills theory, and is still not completely understood. 
Roughly speaking, any affine frame has to be ``soldered'' to the base 
manifold $M$.
Here, soldering means to identify the point $p$ of an affine frame with
a point $x$ of the base manifold $M$. This procedure breaks the translational
part of the original affine invariance and turns it into translation or
diffeomorphism invariance on the base manifold. It is this transition
from internal to external translation invariance which, in the gauge approach,
generates gravity and should deserve future investigations. This should happen
not only on a geometric level but also in the context of Higgs fields and 
their role as generators of mass.

After the soldering procedure one can replace 
the affine gauge connection $(\G^{(T)\a}, 
\G_\a{}^\b)$ by a so--called Cartan connection $(\vt^\a,\G_\b{}^\a)$ 
\cite{cart86, koba63}, where the relation between $\G^\a$ and $\vt^\a$ is 
given by \cite{gron97}
\be
\vt^\a\,:=\,\d^\a_i dx^i+\G^{(T)\a}\,.
\ee
The importance of the introduction of $\vt^\a$ is rooted in the fact that 
under affine transformations $\d_{(\ve\!,\ve_\a{}^\b)}$, generated by 
infinitesimal vector fields $\ve=\ve^i \6_i$ (internal translation) and 
parameters $\ve_\a{}^\b$ 
(general linear transformation), it transforms linearly, i.e., $\vt^\a$ is 
{\it internally} translation invariant:
\be
\d_{(\ve\!,\ve_\a{}^\b)}\vt^\a \,= \ve_\b{}^\a \vt^\b\,.
\ee
Therefore, expressing physical fields by means of the, in general anholonomic,
coframe $\vt^\a$, yields, by construction, an internally translation invariant 
theory, which typically represents a gravitation theory. However, we
stress that this 
construction comes {\it after} the soldering procedure. What remains to
be considered are the general linear transformations and the (external) 
translations on the manifold $M$. These symmetry transformations are the
ones which, in MAG, generate the physically meaningful Noether identities.

Translations on a manifold are generated by the flow of vector fields $\ve$.
The effect of such translations on physical fields is measured by
Lie--derivatives. For physical fields which transform trivially under
GL(n,R)--transformations, for example the Maxwell field or a scalar field, 
the Lie--derivative 
can be taken as the commutator of exterior derivative and interior 
product, $l_\ve...=d(\ve \rfloor...)\;+(\ve\rfloor d...)$. Otherwise, 
in the case of 
spinning matter, e.g., it should be replaced by the gauge--covariant 
Lie--derivative
$\L_\ve...=\DG(\ve\rfloor...)\;+(\ve\rfloor \DG...)$, where $\DG$ denotes the 
$GL(n,R)$--covariant exterior derivative. The operator $\L_\ve$,
in contrast to $l_\ve$, allows to translate tensors 
into tensors, i.e. it is, as its name suggests, gauge covariant and thus 
independent of the orientation of linear frames at different points. 
Therefore it is independent of the linear part of the affine
gauge transformations and leads to gauge--covariant Noether identities, a 
property we want to require for a proper translation 
generator\footnote{For a more 
geometric discussion of this point in favor of the gauge--covariant 
Lie--derivative see \cite{traut73, neem79}.}.

Let us sum up: The gauging of $A(n,R)$ and the subsequent soldering 
procedure has supplemented the initial base manifold $M$ with a Cartan 
connection $(\vt^\a,\G_\b{}^\a)$. A metric $g=g_{\a\b}\vt^\a\otimes\vt^\b$ 
on $M$ is not provided by either the gauging or the, somewhat unclear, 
soldering procedure. Its existence has 
to be postulated, a conceptual drawback which has not been resolved, yet. 
The set of field variables to be put in the gauge Lagrangian is then given by 
$(g_{\a\b}, \vt^\a, \G_\b{}^\a)$. The corresponding field strenghts 
nonmetricity, curvature, and torsion are defined by 
\bea 
{\rm nonmetricity}\qquad
Q_{\alpha\beta}&:=&-{\DG}g_{\alpha\beta}\,=\,-dg_{\a\b}+\G_{\a\b}+\G_{\b\a}\,,
      \label{nonmetricity}\\
{\rm torsion}\quad\qquad T^\alpha&:=&{\DG}\vartheta^\alpha=
d\vartheta^\alpha+\Gamma
_\beta{}^\alpha\wedge\vartheta^\beta \,,\label{torsion}\\
{\rm curvature}\qquad R_\alpha{}^\beta &:=&{}^{\prime\prime}
{\DG}
\Gamma_\alpha{}^\beta{}      
^{\;\prime\prime} =
d\Gamma_\alpha{}^\beta-\Gamma_\alpha{}^\gamma\wedge\Gamma_\gamma{}^\beta
\,.\label{curvature}
\eea
Then the general form of an $A(n,R)-$gauge 
invariant first order gauge Lagrangian 
$V$ becomes
\be
V=V(g_{\a\b}, \vt^\a, Q_{\a\b}, T^\a, R_\a{}^\b)\,.
\ee
An $A(n,R)$--gauge invariant matter Lagrangian with matter fields $\psi$ is 
of the form
\be
L_{\rm mat}=L_{\rm mat}(g_{\a\b}, \vt^\a,\psi, \DG\psi)\,,
\ee 
such that a general model of MAG is determined by a Lagrangian 
\be
L\,=\, V(g_{\a\b}, \vt^\a, Q_{\a\b}, T^\a, R_\a{}^\b) + 
   L_{\rm mat}(g_{\a\b}, \vt^\a,\psi, \DG\psi)\,.
\ee
  
\section{The gauge algebra of MAG}
In order to quantize a gauge theory one needs the knowledge of either the
gauge constraints (Hamiltonian formulation) or the structure of the group of 
gauge transformations which leave the action invariant (Lagrangian
formulation). The latter is determined by the gauge algebra. The gauge 
algebra is the main input required by the covariant BRST-antifield formalism.
It can be determined by commuting the gauge transformations of a given
gauge theory: Suppose we begin with an action functional of the 
form $S_0=\int L(\Phi^i, d\Phi^i)$, where the index $i$ numbers the 
field species of the theory. The gauge 
transformations on the fields $\Phi^i$ can be generated by a generating set
\be
\d_\ve\Phi^i=R^i{}_a(\Phi)\ve^a\,,   \label{gtrafos}
\ee
with $\ve^\a$ the spacetime-dependent gauge parameters and 
$R^i{}_a (\Phi)$ the, in general field dependent, generators
of the gauge transformations. The condensed notation used in (\ref{gtrafos}) 
and in the following was first introduced by DeWitt \cite{dewi65}:
A repeated discrete index not only implies a sum over that index but also
an integration over the corresponding spacetime-variable. Thus, 
formula (\ref{gtrafos}) has to be understood as 
\be
\d_\ve\Phi^i(x)=\int dy R^i{}_a(\Phi)(x,y)\ve^a(y)\,.
\ee
The generating set must be chosen such that it contains all the 
information about the Noether identities. That is,  
from the invariance of the action $S_0$ under the gauge transformations
of a generating set follow the Noether identities in the 
form 
\be
 S_{0,i}R^i{}_a=0\,. \label{noether}
\ee
Having determined a generating set, any gauge transformation $\d \Phi^i$ can 
be written in the form 
\be
\d_\ve\Phi^i=\m^c_aR^i{}_c\ve^a+\m^{ij}_a{{\d S_0}\over{\d\Phi^j}}\ve^a\,,
\quad\;\quad \m^{ij}_a=-(-1)^{\e_i\e_j}\m_a^{ji}\,,  \label{gengauge}
\ee
where the coefficients $\m^c_a$ and $\m_a^{ij}$ are arbitrary functions 
which may involve the fields and
$\e_i$ denotes the Grassmann parity of the field $\Phi ^i$.
The transformations of the form 
\be
\d\Phi^i=\m^{ji}{{\d S_0}\over{\d\Phi^i}}\,,\qquad\quad\m^{ji}=-
(-1)^{\e_i \e_j}\m^{ij}\,,
\label{trivtrafos}
\ee
appearing on the right hand side of (\ref{gengauge}), are called trivial
gauge transformations \cite{dewi79}. They leave the action 
$S_0$ invariant, as is easily verified. Since gauge transformations 
form a group the commutator $[\d_1,\d_2]$ of two gauge transformations is 
again a gauge transformation. Hence it can be expressed in the form 
(\ref{gengauge}),
\be
[\d_1,\d_2]\Phi^i=T^c_{ab} R^i{}_c\ve^b_1\ve^a_2
-E^{ij}_{ab}{{\d S_0}\over{\d\Phi^j}}\ve^b_1\ve^a_2\,, \label{gencomm}
\ee
with, possibly field dependent, structure functions $T^c_{ab}$ and
$E^{ij}_{ab}$. To determine the complete gauge algebra one has to
check for higher order structure functions and calculate higher order 
commutators which manifest themselves in (generalized) Jacobi 
identities \cite{vann81, bata83}. The gauge algebra is said to be open 
if $E^{ij}_{ab}\neq 0$. Otherwise it is called closed. A closed gauge 
algebra with constant structure functions $T^c_{ab}$ specializes to a Lie 
algebra. 

Let us now determine the gauge algebra of MAG. The generating set of MAG is
spanned by the covariant Lie--derivative $\L_\ve$ and the generator of linear
transformations $\d_{\ve_\a{}^\b}$. This defines $R^i{}_a$. 
Thus we have to take into account 
the commutator of two translations generated by $\L_{\ve_1}, \L_{\ve_2}$, the 
commutator 
of a translation $\L_{\ve_1}$ and a general linear transformation  
$\d_{\ve_{2\a}{}^\b}$, and the commutator of two general linear
transformations $\d_{\ve_{1 \a}{}^\b},\d_{\ve_{2 \g}{}^\d}$. 
After some algebra we arrive at the commutation relations 
\bea
[\L_{\ve_1},\L_{\ve_2}]&=&\L_{[\ve_1,\ve_2]} + \d_{(\ve_2
\rfloor(\ve_1\rfloor d\G_\a{}^\b) +
 (\ve_1\rfloor\G_\g{}^\b)(\ve_2\rfloor\G_\a{}^\g)-
    (\ve_2\rfloor\G_\g{}^\b)(\ve_1\rfloor\G_\a{}^\g))}\nonumber \\
 &=&\L_{[\ve_1,\ve_2]} + \d_{(\ve_2
\rfloor(\ve_1\rfloor R_\a{}^\b))} \label{gaugealgebra1}\,,\\
\quad[\L_{\ve_1}, \d_{\ve_{2\a}{}^\b}] &=& \d_{\ve_1\rfloor(\DG\ve_{2\a}{}^\b)}
\,,\label{gaugealgebra2}\\
\quad [\d_{\ve_{1 \a}{}^\b} , \d_{\ve_{2 \g}{}^\d} ] 
&=& \d_{\ve_{1 \d}{}^\rho\ve_{2 \l}{}^\d-\ve_{2 \g}{}^\rho\ve_{1 \l}{}^\g}
\,,\label{gaugealgebra3}
\eea
where, as before, the symbol $\rfloor$ denotes the interior product.

We note that the commutator of two local translations does not 
only yield another local translation but also exhibits a general 
linear transformation involving 
the curvature two-form $R_\a{}^\b = d\G_\a{}^\b-\G_\a{}^\g\wedge\G_\g{}^\b$. 
Also the commutator of a local translation and a general linear transformation
yields as result a general linear transformation which depends on the 
connection $\G_\a{}^\b$. Therfore the gauge algebra of MAG
depends on the field variable $\G_\a{}^\b$ and forms no Lie--algebra.

The commutator of two general linear transformations constitutes a subalgebra 
which resembles the familiar gauge algebra of a 
YM-theory: Denote the $n^2$ generators of $GL(n,R)$-transformations as 
$L^\a{}_\b$. The commutation of two such generators defines, 
according to \cite{dewi65}, structure constants 
$C_\l{}^\rho{}^\a{}_\b{}^\g{}_\d$
\be
[L^\a{}_\b, L^\g{}_\d] = C_\l{}^\rho{}^\a{}_\b{}^\g{}_\d \, L^\l{}_\rho \,.
\label{glnrcomm}
\ee
The structure constants are explicitly given by
\be
C_\lambda{}^\rho{}^\a{}_\b{}^\g{}_{\d}=\d^\a_\d\d^\g_\l\d^\rho_\beta
 -\d^\g_\b\d^\rho_\d\d^\a_\l\,,
\ee
such that the commutation relation (\ref{glnrcomm}) can be rewritten as
\be
[L^\a{}_\b, L^\g{}_\d] = \d^\a_\d L^\g{}_\b-\d^\g_\b L^\a{}_\d\,.
\label{glnrcomm2}
\ee
Commutation of two general linear transformations  
$L_1=\ve_{1\a}{}^\b L^\a{}_\b $,
$L_2=\ve_{2\a}{}^\b L^\a{}_\b$ yields a third transformation 
$L_3=\ve_{3\a}{}^\b L^\a{}_\b $, where the parameter $\ve_{3\a}{}^\b$ 
is determined by the relation
\bea
[\ve_{1\a}{}^\b, \ve_{2\g}{}^\d] &=& \ve_{1\a}{}^\b\ve_{2\g}{}^\d
 C_\l{}^\rho{}^\a{}_\b{}^\g_{}\d \nonumber\\
 &=& \ve_{1\d}{}^\rho \ve_{2\l}{}^\d-\ve_{1\l}{}^\g\ve_{2\g}{}^\rho 
 \nonumber \\
 &=& (\ve_1\ve_2-\ve_2\ve_1)_\l{}^\rho \nonumber \\
 &=& \ve_{3\l}{}^\rho \,.
\eea

Now the structure functions $T^c_{ab}$ and $E^{ij}_{ab}$, which were defined in
(\ref{gencomm}), can be deduced: First we notice that the gauge algebra of 
affine transformations closes and thus
\be 
E^{ij}_{ab}\; =\;0\,.
\ee
Next we define the parameter $\ve^a$ of an affine transformation as the pair
\be
\ve^a:= (\ve^1, \ve^2) :=(\ve, \ve_\a{}^\b)\,.
\ee
From (\ref{gaugealgebra1}) -- (\ref{gaugealgebra3}) we find
\bea
T^1_{11} &=& [...,...] = ...\rfloor d ... - ...\rfloor d ...\,,
\label{cstruc1}\\
T^2_{11} &=& (...\rfloor (...\rfloor d\G_\a{}^\b)) + 
      (...\rfloor(...\rfloor(\G_\g{}^\b\wedge \G_\a{}^\g)))\nonumber \\
  &=&(...\rfloor (...\rfloor R_\a{}^\b)) \,,\\
T^1_{12} &=& T^1_{21} = T^1_{22} =0 \,,\\
T^2_{12} &=& -T^2_{21} = ...\rfloor \DG... \label{cstruc3}\,,\\
T^2_{22} &\equiv& T_\lambda{}^\rho{}^\a{}_\b{}^\g{}_\d 
   =  C_\lambda{}^\rho{}^\a{}_\b{}^\g{}_\d \label{cstruc2}\,.
\eea 
The expressions on the right hand sides of (\ref{cstruc1}) and (\ref{cstruc3}) 
look rather cryptic. This is because condensed notation within the 
formalism of exterior calculus is used. In order to make contact with 
less condensed notation, as introduced in Ref.\ \cite{dewi65}, we evaluate the 
commutator $[\ve_1,\ve_2]$, appearing on the right hand side of
(\ref{gaugealgebra1}), as follows:
\bea
[\ve_1,\ve_2] &=& [\ve^i_1\6_i,\ve^j_2\6_j] \nonumber\\
&=& \ve_1^i\6_i\rfloor d\ve_2^j\6_j-\ve_2^j\6_j\rfloor d\ve_1^i\6_i
\nonumber \\
&=&\ve_1^i\ve_{2,l}^j \d_i^l\6_j-\ve_2^j\ve^i_{1,l}\d^l_j\6_i\nonumber\\
&=&(\ve_1^i\ve_{2,i}^j\d_j^k-\ve_2^j\ve_{1,j}^i\d_i^k)\6_k\,.
\eea
Therefore we obtain from (\ref{cstruc1}) the structure function $T^1_{11}$ 
in the alternative form
\be
T^1_{11}\,\equiv\, T^k_{i'j''}\,=\,\d(x-y')\d(x-z'')_{,i}\d^k_j
                                    -\d(x-z'')\d(x-y')_{,j}\d^k_i\,.
\ee
Converting also the remaining components of $T^c_{ab}$ yields the result
\bea
T^2_{11}&\equiv& T_\a{}^\b{}_{i'j''}\nonumber \\&=&\d(x-y')\d(x-z'')
 \G_{j\a}{}^\b{}_{,i} - \d(x-z'')\d(x-y')
 \G_{i\a}{}^\b{}_{,j}\nonumber\\
&& +\d(x-y')\d(x-z'')\G_{i\g}{}^\b\G_{j\a}{}^\g-
 \d(x-z'')\d(x-y')\G_{j\g}{}^\b\G_{i\a}{}^\g\nonumber\\
&=& \d(x-y')\d(x-z'') R_{ij\a}{}^\b \,, \label{curvdef}\\
T^2_{12}&\equiv& T_\rho{}^\lambda{}_{i'}{}^{\a''}{}_{\b''}\nonumber\\ &=&
\d(x-y')(\d(x-z'')_{,i}+\Gamma_{i\g}{}^\d\rho(L_\d{}^\g)\d(x-z''))
\d^\a_{\rho}\d^\lambda_{\b} \,,\\
T^2_{22} &\equiv& T_\lambda{}^\rho{}^{\a'}{}_{\b'}{}^{\g''}{}_{\d''}
\nonumber\\ 
   &=&  C_\lambda{}^\rho{}^\a{}_\b{}^\g{}_\d\d(x-y')\d(x-z'')\,.
\eea
In Equation (\ref{curvdef}) the components of the curvature tensor got
introduced according to
\be
R_{ij\a}{}^\b \,:=\, \G_{j\a}{}^\b{}_{,i} - \G_{i\a}{}^\b{}_{,j}+
\G_{i\g}{}^\b\G_{j\a}{}^\g-\G_{j\g}{}^\b\G_{i\a}{}^\g\,.
\ee

After having established the structure functions $R^i{}_a$, $T_{ab}^c$,
and $E_{ab}^{ij}$, one has to consider in a next step the Jacobi identity
\be 
\sum_{{\rm cyclic}\; {\rm permutations}\; {\rm of}\; 1,2,3}
\bigl[\d_1, [\d_2, \d_3]\bigr]\, \Phi^i \;=\;0 \label{Jacobi}
\ee
in order to check if it produces any non--trivial relations. If we
substitute in place of any affine gauge transformation the sum of an 
infinitesimal translation and a general linear transformation,
$\d_i=\L_{\ve_i}+\d_{\ve_{i\a}{}^\b}$, it is algebraically straightforward 
to show that the Jacobi identity $(\ref{Jacobi})$ is identically satisfied.
No nontrivial relations are produced such that there are no nonzero
higher order structure functions besides $T^c_{ab}$. 

As final step in the investigation of the gauge algebra it 
remains to observe that the $n$ generators of translations $\L_{\ve_i}$ 
together with the $n^2$ generators of general 
linear transformations $\d_{\ve_\a{}^\b}$ are 
independent, i.e., they define {\it irreducible} gauge transformations. 

We thus conclude from this section that MAG constitutes a closed, 
irreducible gauge theory with field dependent structure functions. 

\section{BRST--antifield symmetry of MAG}
The BRST--antifield construction allows to covariantly
identify physical functions which differ by (i) a gauge transformation 
or (ii) a term proportional to the equations of motion \cite{henn92, gomi95}. 
Such functions 
are viewed as physically indistinguishable. An equivalence class of
physically indistinguishable functions is called an observable.
Covariant quantization requires to consider observables rather than
single functions. The BRST--differential $s$ allows to associate observables
to its cohomology.

In the case of a closed, irreducible gauge theory the construction of $s$ can 
shortly be summarized as follows \cite{henn92}:
One first introduces two differentials, the so--called longitudinal exterior 
derivative $d_L$  and the Koszul--Tate differential $\d_{KT}$. The 
longitudinal 
exterior derivative measures the change of physical functions along the
gauge orbits. Applied to a gauge invariant function, for example, the exterior
longitudinal derivative yields zero. Affiliated with the introduction of 
$d_L$ is the introduction of ghostfields $\h^a$, each of which corresponds to 
an infinitesimal gauge parameter $\ve^a$, and a grading called the 
pure ghostnumber. The Koszul--Tate differential 
allows to identify functions which differ by terms 
proportional to the equations of motion. This requires the introduction
of antifields $\Phi_i^*$ and $\h_a^*$, which correspond
to the fields $\Phi^i$ and $\h^a$, and  a grading called the antighostnumber. 
The BRST--differential $s$ is defined as the sum of $d_L$ and $\d_{KT}$ 
modulo terms which make $s$ nilpotent, $s^2=0$. The grading 
associated to $s$ is called the ghostnumber and given by the 
difference of pure ghostnumber and antighostnumber, gh = puregh - antigh.
The objects of this paragraph are summarized in Table 
\ref{gradtab}. The action of $d_L$ and $\d_{KT}$ on the various fields is 
given by Table \ref{acttab}.

\begin{table}[htb]
\bigskip
\begin{center}
\small
\begin{tabular}{||c|c|c|c||c||}\hline\hline
object & antigh & puregh & $\;\;$gh$\;\;$ & parity \\
\hline\hline
&&&&\\
$\Phi^i$ &0&0&0&$\e_{i}$ \\
&&&&\\
\hline
&&&&\\
$\h^a$ &0&1&1&$\e_{a}+1$ \\
&&&&\\
\hline
&&&&\\
$\Phi_i^*$ &1&0&--1& $\e_{i}+1$ \\
&&&&\\
\hline
&&&&\\
$\h_a^*$&2&0&--2&$\e_{i}$ \\
&&&&\\
\hline
&&&&\\
$\d_{KT}$ &--1&0&1&1\\
&&&&\\
\hline
&&&&\\
$d_L$ & 0&1&1&1 \\
&&&&\\
\hline
&&&&\\
$s$ & not applicable& not applicable& 1& 1\\
&&&&\\
\hline\hline
\end{tabular} 
\end{center}
\caption{The different degrees of the main objects in the irreducible
antifield-construction. The different gradings are the antighostnumber 
associated to $\d_{KT}$, the pure ghostnumber associated to $d_L$, and the
ghostnumber defined by gh = puregh -- antigh.}
\label{gradtab}
\bigskip
\end{table}

\begin{table}[htb]
\bigskip
\begin{center}
\small
\begin{tabular}{||c||c|c|}\hline\hline
\qquad\qquad&$\quad\qquad\d_{KT}\quad\qquad$ & $\quad\qquad d_L\quad\qquad$ \\
\hline\hline
&&\\
$\Phi^i$ & $\d_{KT}\Phi^i=0$ & $d_L\Phi^i=R^i{}_a\h^a$ \\
&&\\
\hline
&&\\
$\h^a$ & $\d_{KT}\h^a=0$ & $d_L\h^a=(-1)^{\e_b}{1\over 2}T^a_{bc}\h^c\h^b$ \\
&&\\
\hline
&&\\
$\Phi^*_i$ & $\d_{KT}\Phi_i^*=-(-1)^{\e_i} S_{0,i}$& $
d_L\Phi^*_i=-(-1)^{\e_a\e_i}\Phi^*_j R^j{}_{a,i}\h^a$ \\
&&\\
\hline
&&\\
$\h_a^*$ & $\d_{KT}\h_a^*=(-1)^{\e_a} 
\Phi_i^*R^i{}_a$ & $d_L\h^*_a=\h_c^*T^c_{ab}
\h^b$ \\
&&\\
\hline\hline
\end{tabular} 
\end{center}
\caption{The action of the Koszul-Tate differential and the longitudinal
exterior derivative on the fields involved in the closed, irreducible 
antifield-construction.}
\label{acttab}
\end{table}

With these definitions the observables are 
obtained as the cohomology classes of $s$ at ghostnumber 0: 
\be
H^0(s)=\{{\rm gauge\;\, invariant\;\, functions\;\, on\;\, shell}\}=
\{{\rm observables}\}  \label{fundament}
\ee
The classical BRST--transformations of a closed and irreducible gauge theory 
are explicitly given by \cite{bata83}
\bea
s\Phi^i &=& R^i{}_a\h^a \,,\label{brst1}\\
s\h^a &=& (-1)^{\e_b}{1\over 2}T^a_{bc}\h^c\h^b\,,\\
s\Phi_i^* &=& -(-1)^{\e_i} S_{0,i}-(-1)^{\e_i \e_a}
\Phi^*_jR^j{}_{a,i}\h^a \nonumber\\
&&\qquad\quad\;-(-1)^{(\e_b+\e_i(\e_b+\e_c+1))}{1\over 2}
\h_a^* T^a_{bc,i}\h^c\h^b\,,\\
s\h_a^* &=& (-1)^{\e_a}\Phi^*_i R^i{}_a +\h_c^* T^c_{ab}\h^b\,.\label{brst4}
\eea

In order to apply this formalism to MAG we first have to enlarge
the field algebra of MAG, which is given by $g_{\a\b}$, $\vt^\a$, and 
$\G_\a{}^\b$, by antifields and ghosts. In particular we need
the following additional fields:
\begin{enumerate} 
\item Ghost fields $\h^\a$, $\h_\a{}^\b$ corresponding to the gauge 
parameters $\ve^\a$ and $\ve_\a{}^\b$. The parameter $\ve^\a$ denotes the
component of the vector field $\ve = \ve^\a e_\a$ which generates an
infinitesimal external translation. Correspondingly we also introduce 
the notation $\h:= \h^\a e_\a$.  
\item Antifields $g^{*\a\b}, \vt^*_\a$, and $\G^{*\a}{}_\b$ of 
antighostnumber 1. In particular, it follows from Table \ref{acttab} that 
they have to fulfill
\bea
\d_{KT} g^{*\a\b}\, &=&\, -{{\d S_0}\over{\d g_{\a\b}}}\,, \label{affanti1}\\
\d_{KT} \vt^*_\a\, &=& \, -{{\d S_0}\over{\d \vt^\a}}\,, \\
\d_{KT} \G^{*\a}{}_{\b} &=& -{{\d S_0}\over{\d\G_\a{}^\b}} \,. 
\eea
\item Antifields $\h^*_\a$, $\h^{*\a}{}_\b$ of antighostnumber 2. 
In the BRST--antifield formalism they got, in fact, introduced in order to 
ensure the validity of the Noether identities and to make the antifields 
$g^{*\a\b}, \vt^*_\a$, $\G^{*\a}{}_\b$ $\d_{KT}$--exact \cite{henn92}. The 
corresponding transformation behavior 
$\d_{KT}\h_a^*=(-1)^{\e_a}\Phi_i^* R^i{}_a$, see Table \ref{acttab}, becomes 
explicitly  
\bea
\d_{KT}\h^*_\a &=& g^{*\g\d}\L_{e_\a}g_{\g\d} 
+ \vt^*_\g \wedge \L_{e_\a}\vt^\g  + \G^{*\g}{}_\d \wedge \L_{e_\a}\G_\g{}^\d
  \,, \\
\d_{KT}\h^{*\a}{}_\b &=& +2g^{*\a\g} g_{\b\g} +
 \vt^*_\b \wedge \vt^\a +\G^{*\a}{}_\b \DG 
 \,. \label{affanti5}
\eea
\end{enumerate}

It is straightforward to obtain explicit expressions like this: Take for
example the term $\Phi_i^* R^i{}_a\h^a$, where $R^i{}_a$ is supposed 
to represent the generating set of $A(n,R)$--gauge transformations.
Its explicit form is derived from the transformation behavior of the fields 
$g_{\a\b}$, $\vt^\a$, and $\G_\a{}^\b$ under external translations
and $GL(n,R)$--gauge transformations: 
One simply replaces gauge parameters by ghosts and contracts with the
corresponding antifield. This furnishes successively the contributions
\bea
\Phi^i&\equiv& g_{\a\b}\,\quad\longrightarrow\quad 
\Phi_i^* R^i{}_a\h^a \,\equiv\, g^{*\a\b}(\L_\h g_{\a\b}
+2\h_{(\a\b)}) \,, \\
\Phi^i&\equiv& \vt^\a \;\quad\longrightarrow \quad 
\Phi_i^* R^i{}_a\h^a \,\equiv\, \vt^*_\a(\L_\h\vt^\a+
 \h_\b{}^\a\vt^\b) \,, \\
\Phi^i&\equiv& \G_\a{}^\b\quad\longrightarrow 
\Phi_i^* R^i{}_a\h^a \,\equiv\, \quad \G^{*\a}{}_{\b}
(\L_\h\G_\a{}^\b -\DG\h_\a{}^\b) \,.
\eea
As another example we consider the term  ${1\over 2} \h_a^* T^a_{bc}\h^c\h^b$. 
Its explicit form follows from the gauge structure functions (\ref{cstruc1}) --
(\ref{cstruc2}). The opposite statistics of the ghosts, relative to the gauge 
parameters, simplifies the commutator structure a bit. We obtain
\bea
\h^*_a&\equiv& \h^*\quad\longrightarrow\quad 
{1\over 2} \h_a^* T^a_{bc}\h^c\h^b \,\equiv\, {1\over 2}\h^*[\h,\h]=
 \h^*(\h\rfloor d\h) \,, \\
\h^*_a&\equiv& \h^{*\a}{}_\b \quad\longrightarrow\quad 
{1\over 2} \h_a^* T^a_{bc}\h^c\h^b \,\equiv\,
\h^{*\a}{}_{\b}(\h\rfloor d\h_\a{}^\b) + \h^{*\a}{}_{\b}(\h_\a{}^\g
\h_\g{}^\b) \,. 
\eea

Now we write down the classical BRST(--antifield)--transformations 
of MAG according to the general formulas (\ref{brst1}) -- (\ref{brst4}):
\bea
sg_{\a\b} &=& \L_\h g_{\a\b} + 2\h_{(\a\b)}  \,,\label{cmag1}\\
s\vt^\a &=& \L_\h \vt^\a+\h_\b{}^\a\vt^\b \,,\\
s\G_\a{}^\b &=& \L_\h\G_\a{}^\b -\DG\h_\a{}^\b \,,\\
s\h^\a &=& -\h\rfloor d\h^\a \,,\\
s\h_\a{}^\b &=& -\h\rfloor(\h\rfloor R_\a{}^\b) 
 -\h\rfloor \DG\h_\a{}^\b -\h_\g{}^\b\h_\a{}^\g \,,\\
sg^{*\a\b} &=& -{{\d S_0}\over{\d g_{\a\b}}} -g^{*\a\b} \L_\h   
- g^{*\a\g}\h_\g{}^\b - g^{*\g\b}\h_\g{}^\a\,,\\
s\vt_\a^* &=& -{{\d S_0}\over{\d \vt^\a}} - \vt^*_\a \L_\h
 -\vt^*_\b\h_\a{}^\b \,,\\
s\G^{*\a}{}_\b &=& -{{\d S_0}\over{\d\G_\a{}^\b}} - \G^{*\a}{}_\b \L_\h
 + \G^{*\g}{}_\d C_\g{}^{\d\a}{}_\b{}^\rho{}_\lambda \h_\rho{}^\lambda 
  \nonumber \\
 &&\qquad +\, \h^{*\g}{}_\d 
C_\g{}^{\d\a}{}_\b{}^\rho{}_\lambda\h_\rho{}^\lambda
  \h\rfloor \, +\, \h^{*\a}{}_\b\h\rfloor(\h\rfloor d) \nonumber \\
 &&\qquad + \, 2 \Bigr[\h^{*\g}{}_\b(\h\rfloor\G_\g{}^\a)\h\rfloor
 \, -\, \h^{*\a}{}_\g(\h\rfloor\G_\b{}^\g)\h\rfloor\,\Bigl] \,,\\
s\h_\a^* &=&  g^{*\g\d}\L_{e_\a}g_{\g\d} 
+ \vt^*_\g \wedge \L_{e_\a}\vt^\g  + \G^{*\g}{}_\d \wedge \L_{e_\a}\G_\g{}^\d
\nonumber  \\
 \quad & &\qquad + \h^*_\b(e_\a\rfloor d\h^\b) +\h^{*\g}{}_\d (e_\a\rfloor 
 \DG \h_\g{}^\d) + 2 \h^{*\g}{}_\d e_\a\rfloor (\h\rfloor R_\g{}^\d)  \,, \\
 s\h^{*\a}{}_\b &=&  +2g^{*\a\g} g_{\b\g} +\vt^*_\b 
\wedge \vt^\a + \G^{*\a}{}_\b \DG
 \nonumber \\
\quad & & \qquad + \h^{*\a}{}_\b \h\rfloor \DG +\h^{*\b}{}_\g\h_\a{}^\g \,.
\label{cmag2}
\eea
If matter fields $\psi$ are present we also have to introduce an antifield
$\psi^*$ corresponding to any matter field $\psi$. The BRST--transformations 
of these fields are of the form
\bea
s\psi &=& (\L_\h + \d_{\h_\a{}^\b}) \psi \,, \\ 
s\psi^* &=& - (-1)^{\e_\psi}{{\d S_0}\over{\d\psi}} - 
            {{\d(s\psi)}\over{\d\psi}}\,. \label{cmag3}
\eea  
 
In the antifield formalism the BRST--transformations are generated by means
of the antibracket $(\;,\;)$ and the extended action $S$. That is, the
BRST--transformation $s\FF$ of a functional $\FF=\FF[\Phi^i, \Phi_i^*,\h^a,
\h_a^*]$ are given by
\be
s\FF \,=\,(\FF, S)\,.  \label{brstgen}
\ee
The antibracket is explicitly defined by the requirements
\bea
\Bigl(\Phi^i(x),\Phi^*_j(x')\Bigr)&=&\d^i_j\d(x-x')\,,\quad
{\rm i.e.\; \; } \quad \Bigl(\Phi^i,\Phi_j^*\Bigr)=\d^i_j\,, \label{br1}\\
\Bigl(\h^a(x),\h^*_b(x')\Bigr)&=&\d^a_b\d(x-x')\,,\quad
{\rm i.e.\; \; } \quad \Bigl(\h^a,\h_b^*\Bigr)=\d^a_b\,. \label{br2}
\eea
such that its action on functionals $\FF$, $\GG$ reads\footnote{The indices
$r$ and $l$ denote right and left differentiation, respectively. So far we
used right differentiation without an index.} 
\be
(\FF, \GG)\; =\; {{\d^r\FF}\over{\d\Phi^i}}{{\d^l \GG}\over{\d\Phi_i^*}}\,-\,
{{\d^r \FF}\over{\d\Phi_i^*}}{{\d^l \GG}\over{\d\Phi^i}}
\,+ \,{{\d^r\FF}\over{\d\h^a}}{{\d^l \GG}\over{\d\h^*_a}}\,-\,
{{\d^r \FF}\over{\d\h^*_a}}{{\d^l\GG}\over{\d\h^a}}\,.
\ee 
The nilpotency of the BRST--transformation, $s^2=0$, is equivalent to the
(classical) {\it Master equation}
\be
(S,S)\,\;=\;\,0\,,
\label{master}
\ee
which can also be taken as the starting point of the BRST--antifield 
construction.

In the case of a closed, irreducible gauge theory the BRST--transformations
(\ref{brst1}), (\ref{brst4}) are generated according to (\ref{brstgen}) if
the extended action $S$ takes the form 
\be
S=S_0+\Phi^*_i R^i{}_a \h^a+ (-1)^{\e_b}
{1\over 2}\Phi_a^*T^a_{bc}\h^c\h^b \,. \label{propsol}
\ee
This is easily proven by taking in (\ref{brstgen}) the functional 
$\FF$ successively as $\Phi$, $\Phi_i^*$, $\h^a$, and $\h_a^*$.
Due to the nilpotency of the BRST--transformations the extended action 
$S$ satisfies automatically the Master equation 
(\ref{master}), i.e. the extended action is BRST--invariant. 
It is a proper BRST--invariant extension of the gauge invariant, classical 
action $S_0$.

According to (\ref{propsol}), the BRST--transformations of MAG, 
(\ref{cmag1}) -- (\ref{cmag3}),
are generated by the extended action
\bea
S=S_0 &+&\int \Bigl(g^{*\a\b}(\L_\h g_{\a\b} + 2\h_{(\a\b)})+ 
       \vt^*_\a\wedge (\L_\h\vt^\a+ \h_\b{}^\a\vt^\b) \nonumber\\ 
&&+ \,  \G^{*\a}{}_{\b}\wedge
       (\L_\h\G_\a{}^\b -\DG\h_\a{}^\b) 
       +\psi^*\wedge(\L_\h\psi+\d_{\h_\a{}^\b}\psi)  \nonumber \\ 
&&+\, +\h^*_\a(\h\rfloor d\h^\a) + 
       \h^{*\a}{}_{\b}(\h\rfloor (\h\rfloor R_\a{}^\b) 
      +\h\rfloor \DG\h_\a{}^\b +
       \h_\a{}^\g \h_\g{}^\b)\Bigr)\,.   \label{caffextact}
\eea

\section{Gauge fixing of two--dimensional MAG}
The BRST--invariant action (\ref{propsol}) is not yet suitable to be used in 
a generating functional of the form 
\be
Z[J]\,\sim\,\int[D\Phi D\h D\Phi^* D\h^*]
\exp\Bigl({i\over \hbar}S(\Phi, \h, \Phi^* , \h^*, J)\Bigr)\,
\quad({\rm not}\; {\rm well}\; {\rm defined})\,. 
\label{wrong}
\ee
This is because  differentiating the Master equation yields an unwanted 
set of gauge transformation under which $S$ is invariant \cite{henn92}. Also 
one would like 
to eliminate the antifields before deriving Green's functions in a
perturbative expansion, simply because there exists no satisfying
physical interpretation of the antifields, yet.   

Following \cite{bata81} we can eliminate both the unwanted gauge invariances 
and the antifields by introducing a {\it gauge fixing fermion
$\Psi$} which is defined to be a functional of fields $\Phi^A$.  
With an appropriate  gauge fixing fermion a gauge fixed extended 
action $S_{fix}$ can be reached by the replacement of antifields by
fields according to 
\be
\Phi^*_A\;=-\;{{\d\Psi}\over{\d\Phi^A}}\,.  \label{antireplace}
\ee
Since $\Phi^*_A$ and $\Phi^A$ are of different parity, $\Psi$ must be
of odd parity, i.e. fermionic.

A field $\Phi^A$, i.e., an original field or a ghost field, possesses 
a ghost number 
$n\geq 0$. The corresponding antifield $\Phi^*_A$ is of ghostnumber
$-(n+1)<0$. Thus, according to equation 
(\ref{antireplace}), $\Psi$ must be of ghostnumber
$-1$. Since $\Psi$ is supposed to be a functional of fields only it
is inevitable to introduce auxiliary fields with negative ghostnumber 
(plus their corresponding antifields in order to maintain the symplectic 
structure on $M$). This can be done straightforwardly since it is always 
possible to add within the BRST-formalism {\it cohomologically trivial pairs} 
that do not change the physical content of the theory:
Consider the auxiliary fields $\overline{\h}^a, b^a$ which are defined to 
satisfy
\be
s\overline{\h}^a=b^a\,,\;\quad sb^a =0\,. \label{auxtrafos}
\ee
The field $\overline{\h}^a$ is not an element 
of the kernel ${\rm Ker}(s)$ while the field 
$b^a$ is both an element of ${\rm Ker}(s)$ and the image ${\rm Im}(s)$. Thus 
both fields $\overline{\h}^a$ and $b^a$ are not contained in 
$H(s)={{{\rm Ker}(s)} \over{{\rm Im}(s)}}$ and do not contribute to the
spectrum of observables. 

According to (\ref{brstgen}) one can 
impose the BRST-transformations (\ref{auxtrafos}) by adding
the auxiliary term $S_{aux}=\int\overline{\h}^*_a\wedge b^a$ to the action $S$:
\be
S\, \longrightarrow \, S_{non-min}\, = \, S+S_{aux}\,=\, 
S+\int\overline{\h}^*_a\wedge b^a \,.
\label{nonmini}
\ee
The extended action together with this supplementary term is an example of a 
non-minimal solution of the master equation. 
The antibracket $(\;, \,)={{\d^R}\over{\d\Phi^A}}{{\d^L}\over
{\d\Phi^*_A}}-{{\d^R}\over{\d\Phi^*_A}}{{\d^L}\over{\Phi^A}}$ now also
contains derivatives involving  the field-antifield pairs 
$\overline{\h}^a$, $\overline{\h}^*_a$ and $b^a$, $b^*_a$.

Finally one has to gauge fix the action $S_{non-min}$ by 
actually choosing a gauge fixing fermion $\Psi$. Not all 
choices are meaningful. The trivial choice $\Psi=0$, for example,
sets all antifields to zero and leads back to the classical action $S_0$. 
There are no definite rules how to choose an appropriate gauge fixing fermion. 

In the following we will illustrate gauge fixing of MAG by means of a 
general two--dimensional model. Gauge fixing procedures for
higher dimensional models follow the same pattern but quickly get
algebraically more complicated. 

We start from the extended action
\bea
S=S_0 &+&\int \Bigl(g^{*\a\b}(\L_\h g_{\a\b} + 2\h_{(\a\b)})+ 
       \vt^*_\a\wedge (\L_\h\vt^\a+ \h_\b{}^\a\vt^\b) \nonumber\\ 
&&+ \,  \G^{*\a}{}_{\b}\wedge
       (\L_\h\G_\a{}^\b -\DG\h_\a{}^\b) 
        +\psi^*\wedge(\L_\h\psi+\d_{\h_\a{}^\b}\psi)  \nonumber \\ 
&&\, +\h^*_\a(\h\rfloor d\h^\a) + \h^{*\a}{}_{\b}
        (\h\rfloor (\h\rfloor R_\a{}^\b) 
      +\h\rfloor \DG\h_\a{}^\b +
       \h_\a{}^\g \h_\g{}^\b)\Bigr)\,.  \label{caffextact2}
\eea
The indices $\a$, $\b$ run from 0 to 1. Here and in the following we will use 
$0,1$ to indicate anholonomic indices and $\tau,\;\s$ to indicate holonomic 
indices. A convention like this is necessary in view of partial derivatives or 
frames which could be understood as holonomic $(\6_i, dx^i)$ or anholonomic 
$(\6_\a=e^i{}_\a\6_i\,,\; \vt^\a=e_i{}^\a dx^i$, with, in general, coordinate
dependent tetrad coefficients $e^i{}_\a$, $e_i{}^\a$). 

Next we  introduce the following $4\times 2=8$ auxiliary fields:
\bea
{\rm translations:}&&\quad \overline{\h}_\a\,,\;b_\a\,,\;\overline{\h}^{*\a}
\,,\;b^{*\a}\,,\\
{\rm linear\;transformations:}&&\quad \overline{\h}_{\a}{}^\b\,,\;b_{\a}{}^\b
 \,,\; \overline{\h}^{*\a}{}_\b\,,\;b^{*\a}{}_\b\,.
\eea
We impose the correct BRST--transformation behavior
of these auxiliary fields by adding the auxiliary term
\be
S_{aux}\,=\,\int(\overline{\h}^{*\a} b_\a+{1\over 2}\overline{\h}^{*\a}{}_\b
b_{\a}{}^\b)
\label{magaux2}
\ee 
to the extended action. 

Now we have to think of an appropriate gauge fixing fermion.
In two dimensional MAG we have six gauge parameters, i.e., 
two parameters $\ve_\a$ of translation invariance and four parameters
$\ve_\a{}^\b$ of general linear invariance. We can
use four of these degrees of gauge freedom to fix the coframe $\vt^\a$ to
the conformal gauge 
\be
\vt^0\,=\, d\tau\,,\qquad\vt^1\,=\, d\s\,. \label{conformal}
\ee
However, this does not fix
the coframe to be orthonormal since the metric components are
independent fields of the theory. In two dimensions, the metric tensor
has three independent components. We can use the remaining two 
degrees of gauge freedom to fix the metric to be diagonal with one 
remaining degree of freedom, $g_{01}=g_{10}=0$, 
$g_{00}=-g_{11}=:(1/2)\exp(\rho)=(1/2)\exp(\rho(\tau,\s))$. 
(We put $g_{00}=-g_{11}$ 
since we assume a Minkowskian signature of the metric. The following procedure
works also for Euclidean signature, though.)  
This gauge can be reached if we choose the gauge fixing fermion as 
(We note that the star $*$ to the {\it 
left} of a field denotes the usual Hodge--star operator.)
\bea
\Psi\,=&&\int\Bigl(*\overline{\h}_0{}^0
(\6_\tau\rfloor\vt^1-\6_\s\rfloor\vt^0)
+*\overline{\h}_1{}^1(\6_\tau\rfloor\vt^0-\6_\s\rfloor\vt^1)\nonumber\\
&&\quad +*\overline{\h}_0{}^1(\6_\tau\rfloor\vt^1+\6_\s\rfloor\vt^0)
+*\overline{\h}_1{}^0(\6_\tau\rfloor\vt^0+\6_\s\rfloor\vt^1-2)\nonumber\\
&&\quad +*\overline{\h}_0 g_{01} +  {}^*\overline{\h}_1 (g_{00}+g_{11})
\Bigr)\,.
\label{mag2fermion2}
\eea
We remove the antifields via the rule $\Phi_A^*\,=\,-{{\d\Psi}
\over{\d\Phi^A}}$ and obtain the explicit replacements
\bea
\overline{\h}^{*0}{}_0&=& -*(\6_\tau\rfloor\vt^1-\6_\s\rfloor\vt^0)\,,
\label{mob1}\\
\overline{\h}^{*1}{}_1&=& -*(\6_\tau\rfloor\vt^0-\6_\s\rfloor\vt^1)\,,\\
\overline{\h}^{*0}{}_1&=& -*(\6_\tau\rfloor\vt^1+\6_\s\rfloor\vt^0)\,,\\
\overline{\h}^{*1}{}_0&=& -*(\6_\tau\rfloor\vt^0+\6_\s\rfloor\vt^1-2)
\label{mob2}\,,\\
\overline{\h}^{*0}&=&-*g_{01}\,,\\
\overline{\h}^{*1}&=&-*(g_{00}+g_{11})\,,\label{mob2a}\\
\vt^*_0&=&-\6_\s\rfloor *\overline{\h}_0{}^0+\6_\tau
\rfloor *\overline{\h}_1{}^1
+\6_\s\rfloor *\overline{\h}_{0}{}^1+\6_\tau\rfloor *\overline{\h}_1{}^0\,,
\label{mob3}\\
\vt^*_1&=&+\6_\tau\rfloor *\overline{\h}_0{}^0-\6_\s
\rfloor *\overline{\h}_1{}^1
+\6_\tau\rfloor *\overline{\h}_{0}{}^1+\6_\s\rfloor *\overline{\h}_1{}^0\,,
\label{mob4}\\
g^{*01}&=&- *\overline{\h}_0\label{mob5}\,,\\
g^{*00}&=& g^{*11}\, =\, - *\overline{\h}_1\,,\label{mob6} \\
\G^{*\a}{}_\b&=&\h^{*\a}\,=\,\h^{*\a}{}_\b\,=\,0\,.\label{mob7}
\eea
Within a path integral we can integrate out the auxiliary variables 
$b_\a$, $b_\a{}^\b$ and also $\vt^\a$, $g_{01}$, and
$g_{00}$. This leads to the gauge conditions 
\bea
\6_\tau\rfloor\vt^1&=&\6_\s\rfloor\vt^0\,, \label{gc1}\\
\6_\tau\rfloor\vt^0&=&\6_\s\rfloor\vt^1\,, \\
\6_\tau\rfloor\vt^1&=&-\6_\s\rfloor\vt^0\,, \\
\6_\tau\rfloor\vt^0+\6_\s\rfloor\vt^1&=&2\,,\\
g_{01}&=&0\,,\\
g_{00}&=&-g_{11}\,,\label{gc2}
\eea
and the gauge fixed action reduces to 
\be
S_{fix}\,=\,S_0 + \int\Bigl( 
g^{*\a\b}(\L_\h g_{\a\b} + 2\h_{(\a\b)})+
\vt_\a^*\wedge (\L_\h\vt^\a +\h_\b{}^\a\vt^\b)\Bigr) \,.
\label{magrest2}
\ee
The single terms that appear in the integral of (\ref{magrest2}) turn out 
to be 
\bea
g^{*\a\b}\L_\h g_{\a\b}&=& 0\,,\\
2g^{*\a\b}\h_{(\a\b)}&=& - *\overline{\h}_1\exp(\rho)(\h_0{}^0+\h_1{}^1)
                        - *\overline{\h}_0\exp(\rho)(\h_0{}^1+\h_1{}^0)\,,\\
\vt^*_\a\wedge \L_\h\vt^\a &=&  *\overline{\h}_0{}^0(\6_\s\h^0- \6_\tau\h^1)
 +  *\overline{\h}_1{}^1(\6_\s\h^1- \6_\tau\h^0) \nonumber\\
&& -\, *\overline{\h}_{0}{}^{1}(\6_\tau\h^1+\6_\s\h^0) -
   *\overline{\h}_1{}^0(\6_\tau\h^0+\6_\s\h^1)\,,\nonumber\\
&& +\,
 *\overline{\h}_0{}^0(\h\rfloor\G_1{}^0 - \h\rfloor\G_0{}^1)
 +  *\overline{\h}_1{}^1(\h\rfloor\G_1{}^1- \h\rfloor\G_0{}^0) \nonumber\\
&& -\, *\overline{\h}_{0}{}^{1}(\h\rfloor\G_0{}^1+\h\rfloor
   \G_1{}^0) -    *\overline{\h}_1{}^0(\h\rfloor\G_0{}^0
                   +\h\rfloor\G_1{}^1)\,,\label{noncompact} \\
\vt_\a^*\wedge\h_\b{}^\a\vt^\b &=&  
 -( *\overline{\h}_1{}^1+  *\overline{\h}_1{}^0)\h_0{}^0
 -( *\overline{\h}_0{}^0+  *\overline{\h}_0{}^1)\h_0{}^1\nonumber\\
 && -\,( *\overline{\h}_0{}^0-  *\overline{\h}_0{}^1)\h_1{}^0
 -( *\overline{\h}_1{}^1-  *\overline{\h}_1{}^0)\h_1{}^1\,.
\eea
The expression (\ref{noncompact}) can be written in a more compact way: We
first note that, due to the conformal gauge,
\be
(\L_\h\vt^\a)(\6_i)\,=\,\6_i\h^\a+(\h\rfloor\G_\b{}^\a)\d_i^\b\,=:\,
D_i\h^\a\,,\label{help2}
\ee
An example of (\ref{help2}) is 
$(L_\h\vt^0)(\6_\tau)=\6_\tau\h^0+\h\rfloor\G_0{}^0=D_\tau\h^0$.
With this notation we can write the term (\ref{noncompact}) in the form
\bea
\vt^*_\a\wedge \L_\h\vt^\a &=&  *\overline{\h}_0{}^0(D_\s\h^0- D_\tau\h^1)
 +  *\overline{\h}_1{}^1(D_\s\h^1- D_\tau\h^0) \nonumber\\
&& -\, *\overline{\h}_{0}{}^{1}(D_\tau\h^1+D_\s\h^0) -
   *\overline{\h}_1{}^0(D_\tau\h^0+D_\s\h^1)\,.
\eea
We collect all pieces and obtain the gauge fixed action
\bea
S_{fix}\,=&& S_0+\int\Bigl( *\overline{\h}_0{}^0(D_\s\h^0- D_\tau\h^1)
 +  *\overline{\h}_1{}^1(D_\s\h^1- D_\tau\h^0) \nonumber\\
&& \qquad\quad+\, *\overline{\h}_{0}{}^{1}(D_\tau\h^1+D_\s\h^0) +
   *\overline{\h}_1{}^0(D_\tau\h^0+D_\s\h^1)\nonumber\\
&&\qquad\quad -( *\overline{\h}_1{}^1+  *\overline{\h}_1{}^0+
   *\overline{\h}_1\exp(\rho))\h_0{}^0 -( *\overline{\h}_0{}^0+ 
    *\overline{\h}_0{}^1+ *\overline{\h}_0\exp(\rho))\h_0{}^1\nonumber\\
&&\qquad\quad -\,( *\overline{\h}_0{}^0-  *\overline{\h}_0{}^1
+  *\overline{\h}_0\exp(\rho))\h_1{}^0
 -( *\overline{\h}_1{}^1-  *\overline{\h}_1{}^0+
   *\overline{\h}_1\exp(\rho))\h_1{}^1\Bigr)\,.\nonumber\\
\label{maggaugefix2}
\eea
Integrating out the ghosts $\h_0{}^0$, $\h_0{}^1$, $\h_1{}^0$, and $\h_1{}^1$ 
yields finally
\be
S_{fix}\,= S_0+\exp(\rho)\int\Bigl( *\overline{\h}_0(D_\tau\h^1-D_\s\h^0)
 + *\overline{\h}_1 (D_\tau\h^0-D_\s\h^1)\Bigl)\,. \label{redux4}
\ee 
This is a general result which does not refer to any particular form of the
initial action $S_0$.

We see that in (\ref{redux4}) the ghosts not only
couple to the conformal factor $\exp(\rho)$ but, via (\ref{help2}),
also to the connection $\G_\a{}^\b$. This feature could
have been expected by the use of the {\it gauge--covariant} Lie--derivative 
as generator of translations. It makes it no longer possible to 
straightforwardly quantize a model based on (\ref{redux4}). The coupling to 
the connection forbids to write down plane--wave solutions for the 
(anti--)ghosts.

As a particular example of (\ref{redux4}) we can consider 
the action of the bosonic string \cite{gree87}  
which is given by the two--dimensional integral 
\be
S_0[X_\m, g_{\a\b}, \vt^\a] =-{1\over 4}\int dX_\m\wedge  *dX^\m=
                  -{1\over 4}\int_{\tau = -\infty}^{\tau=+\infty}
      \int_{\s=0}^{\s=\pi} dX_\m(\tau, \s)\wedge  *dX^\m(\tau,\s)\,.
\label{stringact}
\ee
The integration area, commonly called the {\it world--sheet}, is parameterized 
by the timelike coordinate $\tau$ and the spatial coordinate $\s$. The 
spatial region is supposed to be finite. This is indicated by letting $\s$ 
range from 0 to $\pi$. The fields $X^\mu$ are defined to be scalar fields
on the world--sheet. The index $\mu$ is a priori unrelated to the 
two--dimensional integration area and can be seen as merely numbering the 
scalar fields $X^\mu$. The integrand 
$ -{1\over 4} dX_\m\wedge  *dX^\m$ of the action 
(\ref{stringact}) is not written in a transparent form since it mixes
derivatives of the scalar fields with the integration measure.
We can write more explicitly
\bea
-{1\over 4}dX_\mu\wedge *dX^\mu  &=& -{1\over 4}\6_\a X_\m \6_\b X^\m
\vt^\a\wedge *\vt^\b \nonumber \\
&=& -{1\over 4}\6_\a X_\m \6_\b X^\m g^{\b\d}\h_{\d\g}\vt^\a\wedge\vt^\g
\nonumber \\
&=& -{1\over 2}\6_\a X_\m\6_\b X^\m g^{\a\b}\h\,, \label{moreex}
\eea
where we introduced the two--dimensional volume element 
$\h={1\over 2}\h_{\a\b}\vt^\a\wedge\vt^\b$ with $\h_{\a\b}=
\sqrt{|\det g|}\e_{\a\b}$, $\e_{01}=1, \e_{\a\b}=-\e_{\b\a}$. 
The absence of a connection $\G_\a{}^\b$ 
in (\ref{stringact}) reduces the general formula (\ref{redux4}) to the
well-known quantum action of the bosonic string in conformal gauge, 
\bea
S_{fix}
&=& \int\Bigl(-{1\over 2}(\6_\tau X_\mu\6_\tau X^\mu-\6_\s X_\mu \6_\s X^\mu)
\h \\
&&\qquad + \exp(\rho)( *\overline{\h}_0(\6_\s\h^0 -\6_\tau\h^1)
 +  *\overline{\h}_1(\6_\s\h^1- \6_\tau\h^0))\Bigr).\label{redux1}
\eea
Now the ghost are only coupled to the conformal factor $\exp(\rho)$ 
which is the origin of the conformal anomaly. 

In the context of string theory, models based on 
(\ref{redux4}) with nontrivial connection $\G_\a{}^\b$ seem to have not 
been investigated, yet. Also the quantization of some other 
two--dimensional model of MAG, based on the general solution (\ref{redux4}), 
seems to have never been conducted.

\section*{Acknowledgment}
The author would like to thank Professor F.W.\ Hehl for carefully reading the
manuscript, useful advice, and constant support

\end{document}